\documentclass[twocolumn]{aastex63}
\usepackage{graphicx}
\usepackage{blindtext}
\usepackage{amsmath}
\usepackage{mathtools}
\usepackage{multirow}
\usepackage{hyperref}
\usepackage{makecell}
\usepackage{tabu}
\hypersetup{
	colorlinks	= true,
	linkcolor	= red,
	urlcolor	= cyan,
	citecolor	= blue
}
\usepackage{longtable}
\usepackage{lineno}

\newcommand{\exoclock}{\mbox{ExoClock}}
\newcommand{\ariel}{\mbox{\textit{Ariel}}}

\begin{document}

%\linenumbers

% TITLE
\title{\exoclock\ project II: A large-scale integrated study with 180 updated exoplanet ephemerides}

\correspondingauthor{A. Kokori}
\email{anastasia.kokori.19@ucl.ac.uk}

\author{A. Kokori}
\affiliation{University College London, Gower Street, London, WC1E 6BT, UK}

\author{A. Tsiaras}
\affiliation{University College London, Gower Street, London, WC1E 6BT, UK}

\author{B. Edwards}
\affiliation{University College London, Gower Street, London, WC1E 6BT, UK}

\author{M. Rocchetto}
\affiliation{University College London, Gower Street, London, WC1E 6BT, UK}

\author{G. Tinetti}
\affiliation{University College London, Gower Street, London, WC1E 6BT, UK}

\author{L. Bewersdorff}
\affiliation{Amateur Astronomer\footnote{A list of associated private observatories contributed to this work can be found in Appendix A}}

\author{Y. Jongen}
\affiliation{Observatoire de Vaison-La-Romaine, Départementale 51, près du Centre Equestre au Palis - 84110 Vaison-La-Romaine, France}

\author{G. Lekkas}
\affiliation{Department of Physics, University of Ioannina, Ioannina, 45110, Greece}

\author{G. Pantelidou}
\affiliation{Department of Physics, Aristotle University of Thessaloniki, University Campus, Thessaloniki, 54124, Greece}

\author{E. Poultourtzidis}
\affiliation{Department of Physics, Aristotle University of Thessaloniki, University Campus, Thessaloniki, 54124, Greece}

% ...

\begin{abstract}

The \exoclock\ project is an inclusive, integrated, and interactive platform that was developed to monitor the ephemerides of the \ariel\ targets to increase the mission efficiency. The project makes the best use of all available resources, i.e., observations from ground telescopes, mid-time values from the literature and finally, observations from space instruments. Currently, the \exoclock\ network includes 280 participants with telescopes capable of observing 85\% of the currently known \ariel\ candidate targets. This work includes the results of $\sim$1600 observations obtained up to the 31st of December 2020 from the \exoclock\ network. These data in combination with $\sim$2350 mid-time values collected from the literature are used to update the ephemerides of 180 planets. The analysis shows that 40\% of the updated ephemerides will have an impact on future scheduling as either they have a significantly improved precision, or they have revealed biases in the old ephemerides. With the new observations, the observing coverage and rate for half of the planets in the sample has been doubled or more. Finally, from a population perspective, we identify that the differences in the 2028 predictions between the old and the new ephemerides have an STD that is double what is expected from gaussian uncertainties. These findings have implications for planning future observations, where we will need to account for drifts potentially greater than the prediction uncertainties. The updated ephemerides are open and accessible to the wider exoplanet community both from our Open Science Framework (OSF) repository and our website.

\end{abstract}

% TITLE

\section{INTRODUCTION} 

Follow-up observations of known transiting exoplanets are important to properly plan future observations with larger facilities in order to avoid wasting part of their observing time. While the transit times of near future events are predicted with relatively good accuracy, several factors prevent accurate predictions far into the future. Firstly, the limited amount of available data for each planet introduces biases on the ephemeris estimation \citep[e.g][]{Benneke2017, 2019AnA...622A..81M}. Moreover, the precision of the predicted transit times is degrading with time due to the uncertainties of the initial ephemerides \citep[e.g][]{2019AnA...622A..81M}. For example, planets that were recently discovered by TESS, had initial ephemerides with very high uncertainties \citep[e.g][]{Dragomir2020, Zellem2020}. These ephemerides were only improved by follow-up observations and data from the extended TESS mission. Other issues that might result in biases in transit times include tidal orbital decays, gravitational interactions with other bodies or apsidal precession \citep[e.g][]{Agol2005, Maciejewski2016, Bouma2019}.

Future space missions aiming to characterise exoplanets require a good knowledge of transit times in order to increase the efficiency of the mission. The \ariel\ mission will spectroscopically observe the atmospheres of 1000 planets in order to investigate their nature. It will observe thousands of transits so it is crucial to improve the currently known ephemerides. 

The importance and efficiency of using small sized telescopes for observing transits has been highlighted in several works \citep[e.g][]{Beck2019, Kabath2019, 2019AnA...622A..81M, Kokori2021, 2021MNRAS.504.5671E, Zellem2020}. In this regard, their contribution to planning observations for future space missions is of high significance. Since \ariel\ will observe a large number of planets, it is necessary to provide a list with verified ephemerides before the launch of the mission.

The best use of resources can be achieved through large scale efforts which integrate data from various sources. These sources include data from the literature, data from telescopes of any size -- even small ones -- and, finally, data from space for the most challenging targets. \exoclock\ is such a project; it is an open, integrated platform that aims to monitor the ephemerides of the \ariel\ candidate targets. The nature, architecture and organisation of the project are described in detail in \cite{Kokori2021}.

In \cite{Kokori2021} a first round of ephemerides updates was presented for 28 planets. The \exoclock\ project has been in operation since September 2019 and in the course of this effort it became clear that many planets have large scatter in the O-C diagrams or they don’t have many observations. Therefore, such targets require continuous monitoring and a longer time coverage of observations to decrease their current or predicted uncertainties.

In this work, we present the first large-scale update to the ephemerides of 180 planets. The refined ephemerides have been derived from a combination of mid-times from both \exoclock\ observations and literature data. The literature research includes a collection of the majority of mid-times from previous publications for all the 180 planets. The results show that a considerable number of planets have mid-times derived from only a small number of past observations, in many cases from the discovery data alone. The absence of an adequate number of observations leads to increased uncertainties in the ephemerides. This highlights the importance of the continuous monitoring provided by \exoclock. Apart from its role to support the \ariel\ space mission, \exoclock\ aims to act as a community service with reliable ephemerides to be utilised for other exoplanet research purposes.

\section{The \exoclock\ network} 

At the time of writing, the \exoclock\ network includes 280 participants (80\% are amateur astronomers) and 300 telescopes with sizes ranging between 6 and 40 inches (80\% are smaller than 17 inches). We calculate the S/N for each planet-telescope combination based on Equation \ref{eq:snr}. If S/N is higher than 15, then the planet is flagged as accessible.

\begin{equation}
\label{eq:snr}
S/N = a D \sqrt{10 ^ {\frac{12 - R_{mag}}{2.5}}} \frac{T_{Dp}}{ \sqrt{ \frac{1}{T_{Dr}} + \frac{1}{120}} } 
\end{equation}

where $a$ is a constant, $D$ is the telescope aperture in inches, $R_{mag}$ is the R magnitude of the star, $T_{Dp}$ is the transit depth in mmag, and $T_{Dr}$ is the transit duration in minutes.

\begin{figure}
\centering
\includegraphics[width=\columnwidth]{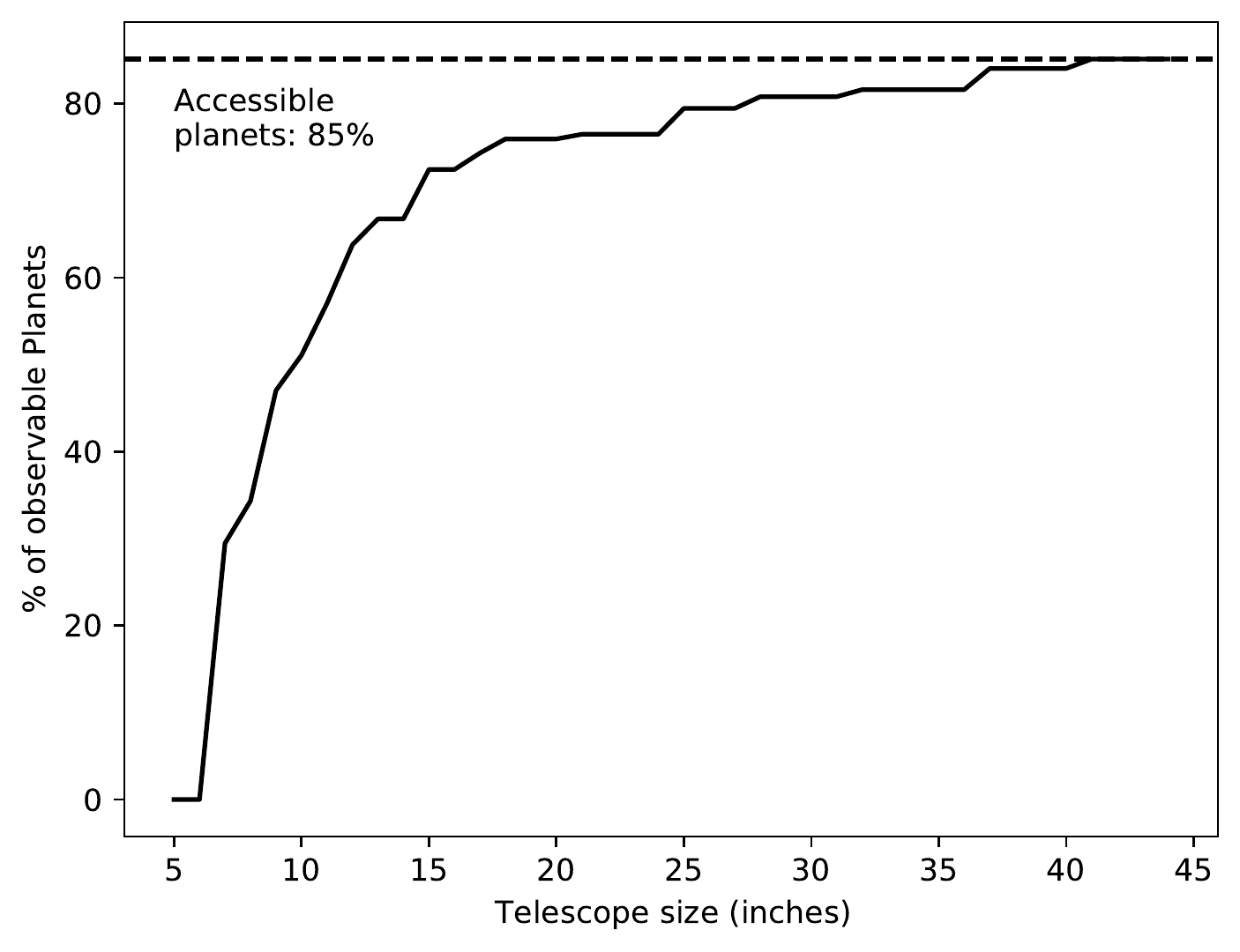}
\caption{Cumulative percentage of observable planets as a function of the telescope size.}
\label{fig:network_capabilities}
\end{figure}

The factor $a$ was estimated empirically based on observations acquired from the Holomon Astronomical station of the Aristotle University of Thessaloniki in Greece. The data were obtained with an 11-inch telescope, an ATIK11000 camera and a Red (Cousins) filter. This can be considered as a representative example of a system capable of performing transit observations. Assuming that an observation starts one hour before the transit, ends one hour after the transit, has an exposure time of one minute, and has overheads of 30 seconds, the value of $a$ is equal to 0.0125. In the course of the project, this factor is updated for each telescope based on the quality of the observations acquired. 

Compared to \cite{Kokori2021}, here we can take into account the updated number and the updated capabilities of the telescopes in our network -- alongside the observability constraints (host star above 20 degrees, transit duration shorter than 6 hours) -- and better estimate the capabilities of the network as a whole. Figure \ref{fig:network_capabilities} shows the cumulative percentage of observable planets as a function of the telescope size. The large number and the global distribution of telescopes smaller than 17 inches ensures that the majority of our targets (75\%) are observable with this type of equipment. The larger telescopes can contribute an additional 10\%, leading to a total of 85\% of observable planets from the list of currently known planets that are candidates for \ariel\ \citep{Edwards2019}. Given the limited time available to larger facilities, this kind of telescope distribution is proven to be very efficient for achieving a large-scale follow-up program.

\section{Data acquisition and evaluation} 

\subsection{\exoclock\ data} \label{sec:exoclock_data}

As part of the \exoclock\ website\footnote{\href{https://www.exoclock.space}{exoclock.space}} we provide a personalised scheduler for each observer, a suggested protocol on how to acquire data and the data analysis software\footnote{\href{https://github.com/ExoWorldsSpies/hops}{github.com/ExoWorldsSpies/hops}} to perform reduction and photometry. We refer the interested reader to \cite{Kokori2021} for more details. The most important aspects of the observing protocol include the use of the Red Cousins filter, the acquisition of data from one hour before the transit start until one hour after the transit end, and the regular check of the system clock to ensure accurate time-stamping of data. The above do not constitute requirements but only suggestions, as we welcome contributions of any kind. If the uploaded data do not include uncertainties, these are estimated with a moving standard deviation.

To ensure the high quality and homogeneity of observations, we perform light curve modelling on the \exoclock\ website, using our dedicated exoplanet catalogue for the planet parameters (the Exoplanet Characterisation Catalogue) and the open source Python package PyLightcurve\footnote{\href{https://github.com/ucl-exoplanets/pylightcurve}{github.com/ucl-exoplanets/pylightcurve}} \citep{Tsiaras2016B2016ascl.soft12018T} for transit modelling. Again, we refer the interested reader to \cite{Kokori2021}, while here we give a summary of the process. For every transit, we convert the time to BJD$_\mathrm{TDB}$, we then fix all the transit parameters with the exception of the planet-to-star radius ratio and the mid-time, and also we calculate the limb-darkening coefficients using the ExoTETHyS\footnote{\href{https://github.com/ucl-exoplanets/ExoTETHyS}{github.com/ucl-exoplanets/ExoTETHyS}} \citep{Morello2020} package for the specific photometric filter used. Finally, we fit the light curve with a transit model (exposure-integrated) together with a trend model (linear with airmass, linear with time, or quadratic with time based on the chi-squared of the residuals) using the nested sampling techniques as implemented in the Nestle package\footnote{\href{https://github.com/kbarbary/nestle}{github.com/kbarbary/nestle}}. After a first fit on the data, we scale the uncertainties to match the standard deviation of the residuals and re-fit. This way, we take into account any extra scatter in the observation and end up with a conservative estimate of the uncertainties in the final results.

\begin{figure}
\centering
\includegraphics[width=\columnwidth]{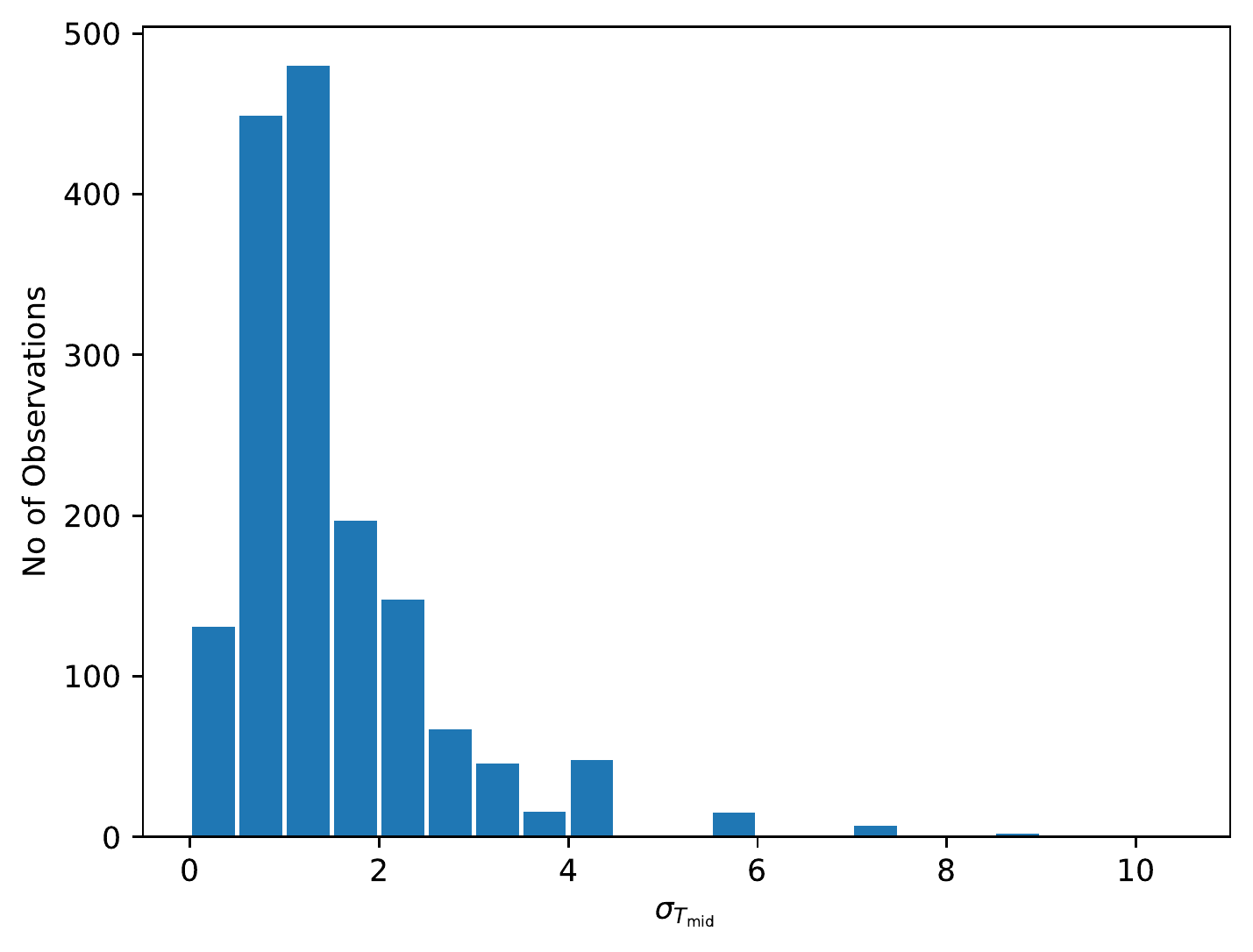}
\caption{Distributions of the achieved uncertainties on the mid-transit time from all the \exoclock\ observations.}
\label{fig:oc_errors}
\end{figure}

In this work, we considered 180 planets that had at least two \exoclock\ observations submitted to the website before the end of 2020 and which met the required quality standards. We assess the quality of each light-curve individually, based on three criteria, as listed below.

\begin{figure*}
\centering
\includegraphics[width=\textwidth]{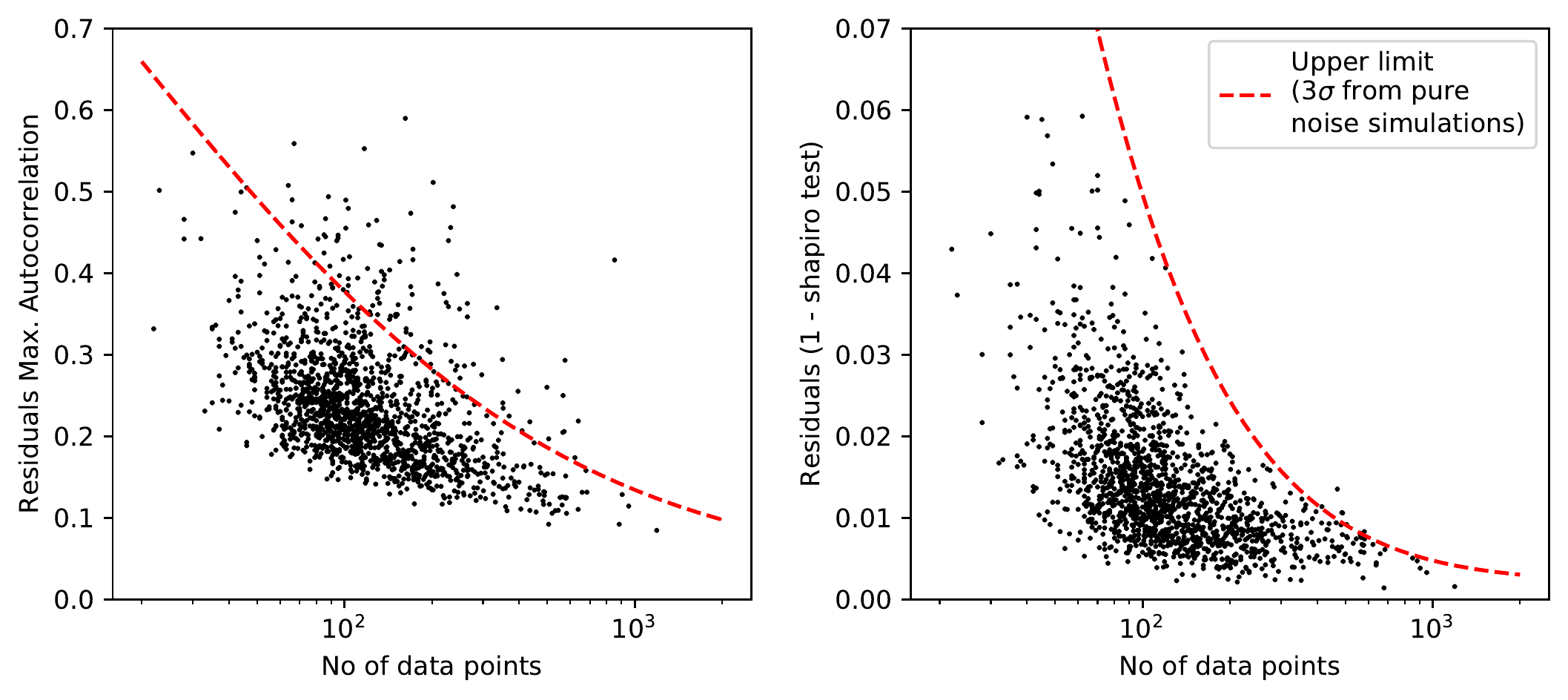}
\caption{Residuals autocorrelation (left) and Shapiro (right) tests for all the \exoclock\ observations as a function of the number of data points in each light-curve. The red dashed line indicates the upper acceptable limit for each metric, as calculated from time series of white noise.}
\label{fig:autocorrelation_shapiro}
\end{figure*}

\begin{enumerate}
\item The fitted $R_p/R_s$ should not differ by more than 3$\sigma$ from the expected literature value. The only exceptions to this rule are planets that orbit stars with physical or projected companions (e.g. HAT-P-1b, K2-29b, WASP-77Ab, WASP-85Ab, XO2-Nb) or planets with grazing transits (e.g. WASP-45b, WASP-67b).

\item The autocorrelation and the Shapiro statistic for the normalised residuals should not differ by more than 3$\sigma$ from the values for pure white noise at the same time. We estimated these limits as a function of the number of data points in a light-curve from 100,000 simulated time series. The advantage of these metrics is that they are very sensitive to systematic noise and to outliers.

\item The fitted transit time should not have an uncertainty greater than 10 minutes and it should be in agreement with other observations on the same day or a few days apart (if such observations exist). 
\end{enumerate}

The total number of approved light-curves for this data release is $\sim$1600, spanning the period from 2008 to 2020. For the majority ($\sim$60\%) of the planets, the number of observations is lower than or equal to 10, indicating that more observations in the future are necessary. (For a discussion on the coverage achieved by our observations see Section \ref{sec:exoclock_contribution}).

The overall precision achieved in the transit timing can be evaluated from Figure \ref{fig:oc_errors}, where the majority of the observations ($\sim$~90\%) deliver a precision better than three minutes. Additionally, Figure \ref{fig:autocorrelation_shapiro} shows the autocorrelation and the Shapiro diagnostics as a function of the light-curve length, together with the acceptable upper limit. Of these light-curves, 7\% fail the autocorrelation test mostly due to spot-crossing events or weather. However, these light-curves pass the $R_p/R_s$ and the Shapiro test as these events do not cause strong residuals. These light-curves will be flagged in the final data product to warn anyone who wishes to use them.

\subsection{Literature data}

Approximately 2350 transit mid-time values were collected from the literature for the 180 planets in our sample. While collecting the literature mid-times we:

\begin{enumerate}
\item included individual transit time measurements and not ephemerides, with the exception of the discovery paper;

\item included transit time measurements with uncertainties less than 10 minutes (similarly to the restriction for the \exoclock\ data);

\item did not include measurements that came from the Exoplanet Transit Database \citep[ETD,][]{Poddany2010} to avoid duplications, as we have started sharing data directly with the ETD collaboration (see \ref{sec:etd}) and these data will be directly linked to \exoclock;

\item converted the reported time formats to BJD$_\mathrm{TDB}$.

\end{enumerate}

We need to note here that 35\% of the planets in our sample have only one data point from the literature, related to the discovery of the planet. This statistic reveals a significant gap in the literature and highlights the need for a large-scale follow-up project like \exoclock.

\subsection{ETD data} \label{sec:etd}

In an effort to make the best use of all available resources, we would like to collaborate with other ground networks which have current or past observations of the \ariel\ candidate targets. At the moment the largest database of such observations is the Exoplanet Transit Database \citep[ETD,][]{Poddany2010} run by the Czech Astronomical Society since 2009, which provides more than 10,000 transit light-curves for more than 350 exoplanets systems. In this study, we included 18 observations for three planets provided by the ETD network and mid-times from these were integrated with the \exoclock\ and the literature data.

In order to maintain homogeneity and reliability in our analysis, we only considered observations with a data quality index lower than three which were then processed on the \exoclock\ website using the same methodology and validation criteria as for the \exoclock\ data (Section \ref{sec:exoclock_data})

Despite the low number of observations used in this analysis, we value the contribution of ETD and this is the first collaborative work between the two networks. Such collaborations are critical to avoid duplications and waste of resources. We aim to continue our collaboration and gradually integrate more data from ETD in future publications.

\section{Results} \label{sec:results}

\subsection{Updated ephemerides for 180 planets}

Here we present updated ephemerides for 180 of the total of 370 planets that are currently in the \exoclock\ target list. To determine these ephemerides, we combined all the available data. First, we updated the zero epoch to the weighted average of the available epochs, and then fitted a line on the epoch vs mid-transit times data. To do so we used the MCMC algorithm as implemented in the emcee package \citep{ForemanMackey2013}. After a first fit, we scaled-up the uncertainties so that the mean uncertainty was equal to the STD of the O-C residuals and in this way incorporate any non-gaussian noise. While, this step did not have a significant effect on the values of the zero-epoch mid-transit and the period, it was important for their uncertainties. Without scaling, the uncertainties on the final ephemerides would had been largely underestimated, leading to reduced chi-squared values larger than 1. Table \ref{tab:updated_ephemerides} in Appendix \ref{sec:updated_ephemerides} provides all the new ephemerides and references to the literature values used.

\begin{figure}
\centering
\includegraphics[width=\columnwidth]{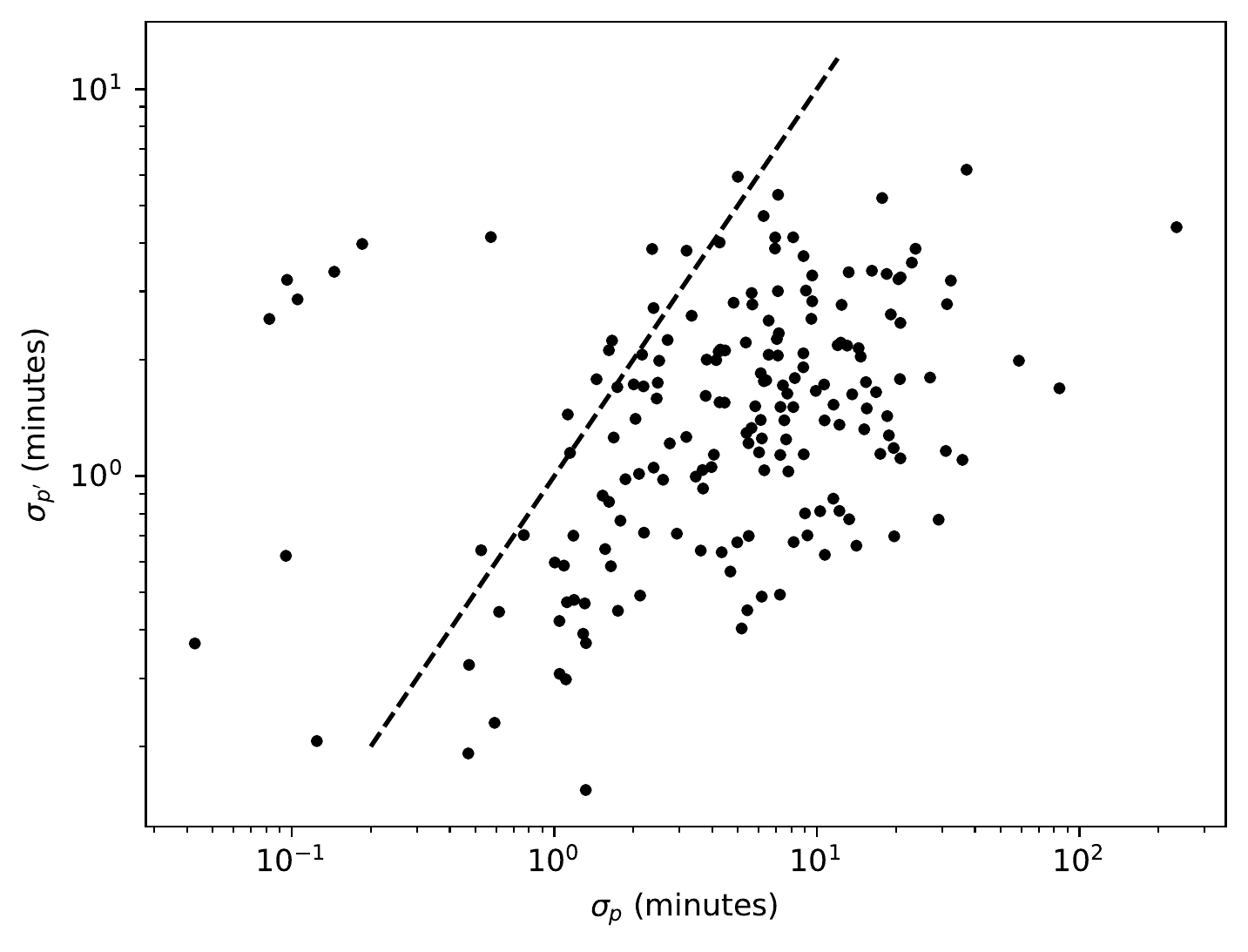}
\caption{Uncertainties for the transit time prediction at the end of 2028 as calculated with the old ephemerides (horizontal axis) and with the new ephemerides as estimated from \exoclock\ (vertical axis). The dashed line is the line of equal values, and the planets for which the prediction precision has been improved (90\%) are plotted on the right of this line.}
\label{fig:before_after}
\end{figure}

\begin{figure}
\centering
\includegraphics[width=\columnwidth]{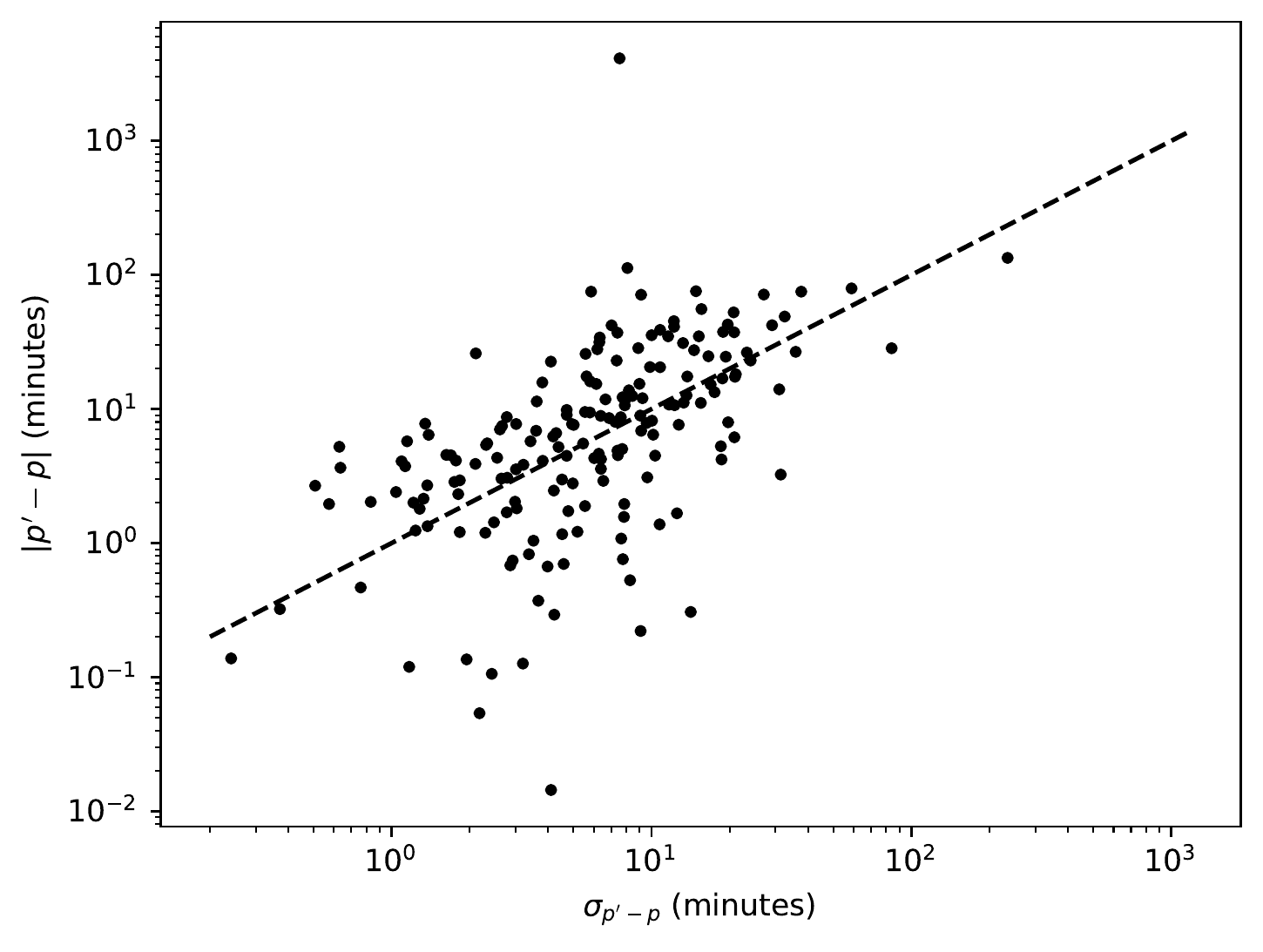}
\caption{Difference in the transit time prediction at the end of 2028 between the old and the new ephemerides, as a function of its uncertainty (quadratically combined). The dashed line is the line of equal values, and the planets for which the drift is greater than 1$\sigma$ (57\%) are plotted above this line.}
\label{fig:drifts}
\end{figure}

Figure \ref{fig:before_after} shows the uncertainties in the 2028 predictions before and after the updates presented in this work ($\sigma_p$ and $\sigma_{p'}$, respectively). We need to note that all the new predictions have uncertainties lower than 10 minutes and an improvement has been achieved for 162 (90\%) of them. There is a small number of planets (6) for which the prediction uncertainty has increased from $\sim$0.1 minute to $\sim$1-4 minutes. These planets were observed by Kepler/K2 but the individual mid-time data were not reported in the literature, hence only the initial ephemerides were used. We plan to re-analyse and add the individual Kepler/K2 light curves in our database in our future data releases, solving the above issue. Moreover, Figure \ref{fig:drifts} shows the difference in the 2028 predictions between the new and the old ephemerides as a function of uncertainty, where 103 planets (57\%) have drifts greater than their uncertainties.

\subsection{Deviations from linear ephemerides}

To assess the gaussianity of the final O-C diagrams we use the same methodology as for the individual light-curves: by evaluating the autocorrelation and the Shapiro test. 

From the autocorrelation test we find that three planets -- WASP-12b, WASP-10b, and Qatar-2b -- have a statistically significant O-C autocorrelation ($>3\sigma$ deviation from Gaussian noise) that indicates a non-linear ephemeris. WASP-12b is known to have a non-linear ephemeris as its orbit is decaying \citep{Maciejewski2016, Turner2021}, while for the other two planets more observations are required to verify the non-linear nature of their ephemerides.

Additionally, the Shapiro test can help us identify high frequency noise in O-C diagrams. We find five planets that deviate significantly from pure noise based on the Shapiro test: GJ1214b, HAT-P-32b, TrES-3b, WASP-43b, WASP-79b. For the planets mentioned above, an additional noise component should be taken into account when predicting future transit events, and we have indicated this in the Catalogue of \exoclock\ Ephemerides (Section \ref{sec:catalogue_of_ephemrides}). As far as high frequency noise is concerned, we find the most significant deviation (15 times higher than the 3$\sigma$ limit) is for TrES-3b, a planet for which there are contradicting results in the literature regarding the presence of an additional planet or not \citep[e.g.][]{2020AJ....160...47M}.

\section{Data release B}

The second data release of the \exoclock\ project includes two data products: the catalogue of \exoclock\ observations and the catalogue of \exoclock\ ephemerides. All data products and their descriptions can be found through the OSF repository with DOI: \href{http://doi.org/10.17605/OSF.IO/WNA5E}{10.17605/OSF.IO/WNA5E}.

\subsection{Catalogue of \exoclock\ Observations}

The new catalogue of \exoclock\ observations contains $\sim$1600 light-curves analysed in this work, which refer to 180 planets from the \exoclock\ target list. These observations were conducted between 2008 and 2020, submitted to the \exoclock\ platform before the end of 2020, and validated according to the criteria described in Section \ref{sec:exoclock_data}. From the $\sim$1600 observations in this dataset, 66\% were obtained by amateur astronomers and the remaining 34\% by professionals.

In the online repository, each observation is accompanied by:
\begin{enumerate}
\item metadata regarding the observer(s), the planet observed (link to ECC), the equipment used (telescope-camera-filter), the exposure time and the time and flux formats;
\item the raw light curve filtered for outliers, converted to BJD$_\mathrm{TDB}$ and flux formats and enhanced with the estimation for the uncertainties, the target altitude, and the airmass;
\item the fitting results, including the de-trending method used and its parameters;
\item the de-trended light curve, enhanced with the de-trending model, the transit model and the residuals;
\item fitting diagnostics on the residuals (standard deviation, chi-squared, autocorrelation).
\end{enumerate}

\subsection{Catalogue of \exoclock\ Ephemerides} \label{sec:catalogue_of_ephemrides}

The new catalogue of \exoclock\ ephemerides contains the updated ephemerides for 180 planets from the \exoclock\ target list (see also Table \ref{tab:updated_ephemerides} in Appendix \ref{sec:updated_ephemerides}). 

In the online repository, each observation is accompanied by:
\begin{enumerate}
\item the mid-time values used to calculate the ephemeris;
\item references for the literature data used;
\item links to the \exoclock\ and ETD data used;
\item flags concerning the detection of non-linear ephemerides or high frequency noise in the O-C diagrams.
\end{enumerate}

\section{Discussion}

\subsection{\ariel\ - focused implications}

\ariel\ will observe 75\% of the transit duration before and after each transit to allow for the correction of instrumental systematics. For this reason, in \cite{Kokori2021} we defined that the aim of the \exoclock\ project is to deliver ephemerides that can predict the transit times for the end of 2028 with a precision higher than 1/12 of the transit duration (target uncertainty).

From the results presented in section \ref{sec:results}, three classes of planets can be distinguished:

\begin{itemize}

\item class 1: 31 planets (17\%) had initial ephemerides with prediction uncertainties greater than the target uncertainties; 

\item class 2: 41 planets (24\%) had initial ephemerides with prediction uncertainties lower than the target uncertainties, but the new ephemerides give predictions that deviate significantly from the initial ones (more than the target uncertainties); 

\item class 3: the remaining 108 planets (60\%) had initial ephemerides with prediction uncertainties lower than the target uncertainties, and the new ephemerides give predictions that do not deviate significantly from the initial ones.

\end{itemize}

In total, 40\% of the ephemerides that have been updated in this work are important for the planning of \ariel\ observations. Despite the fact that the remaining 60\% of the planets had reliable initial ephemerides, the majority of them still have a limited number of observations (see \ref{sec:exoclock_contribution}) and drifts might appear in the future. Therfore, continued monitoring is necessary.

\subsection{The contribution from \exoclock} \label{sec:exoclock_contribution}

The overall contribution of the current work can be assessed by examining the distribution of the observations over the years for each planet. To do so, we used the following metrics:

\begin{itemize}
\item mean observation rate: total number of data points divided by years since the first data point.
\item observing coverage: number of years with at least one data point divided by years since the first data point.
\end{itemize}

\begin{figure}
\centering
\includegraphics[width=\columnwidth]{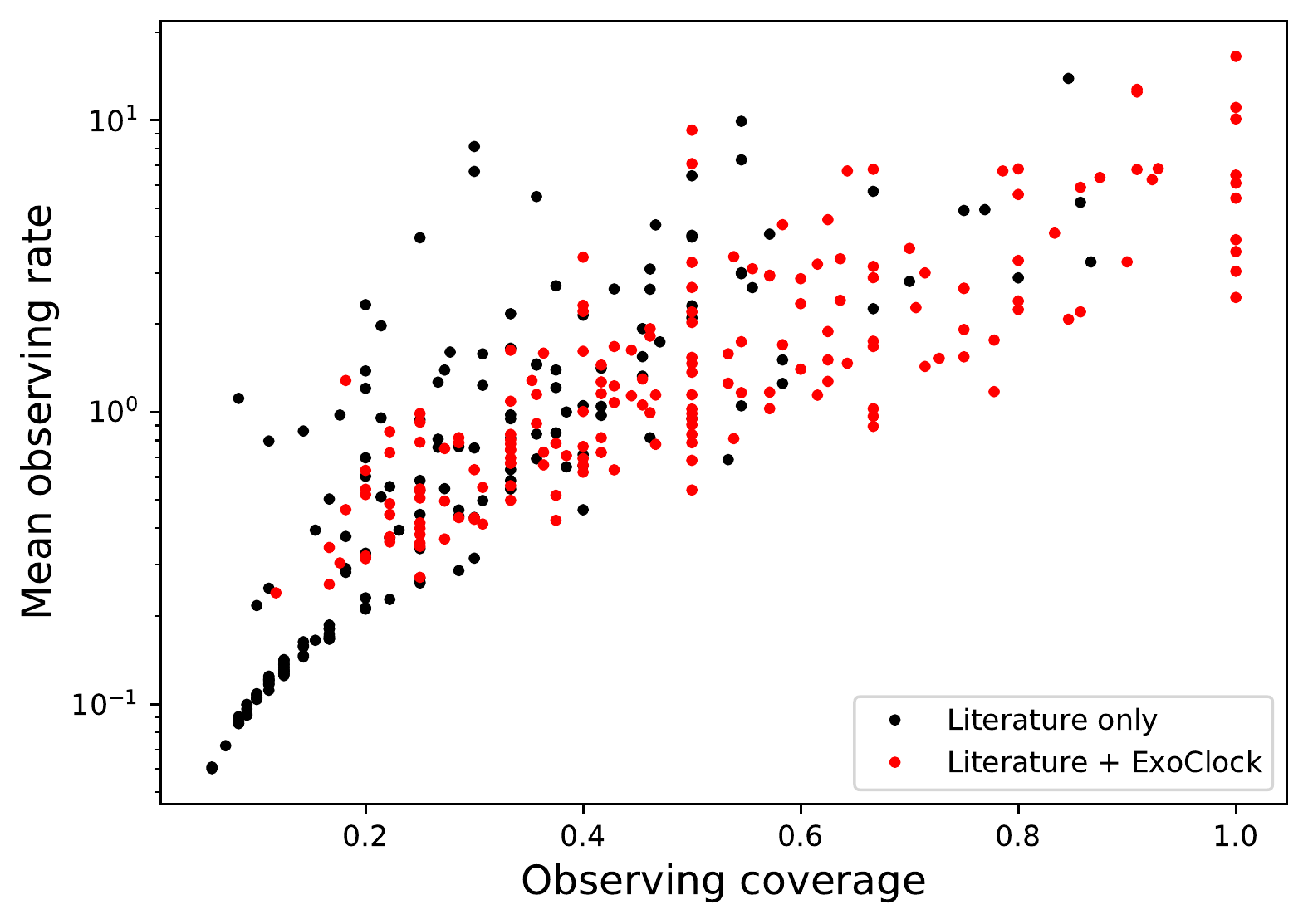}
\caption{Comparison of the mean observing rate vs observing coverage collected from literature data and from the combination of both literature and \exoclock\ data.}
\label{fig:rate_vs_coverage}
\end{figure}

\begin{figure}
\centering
\includegraphics[width=\columnwidth]{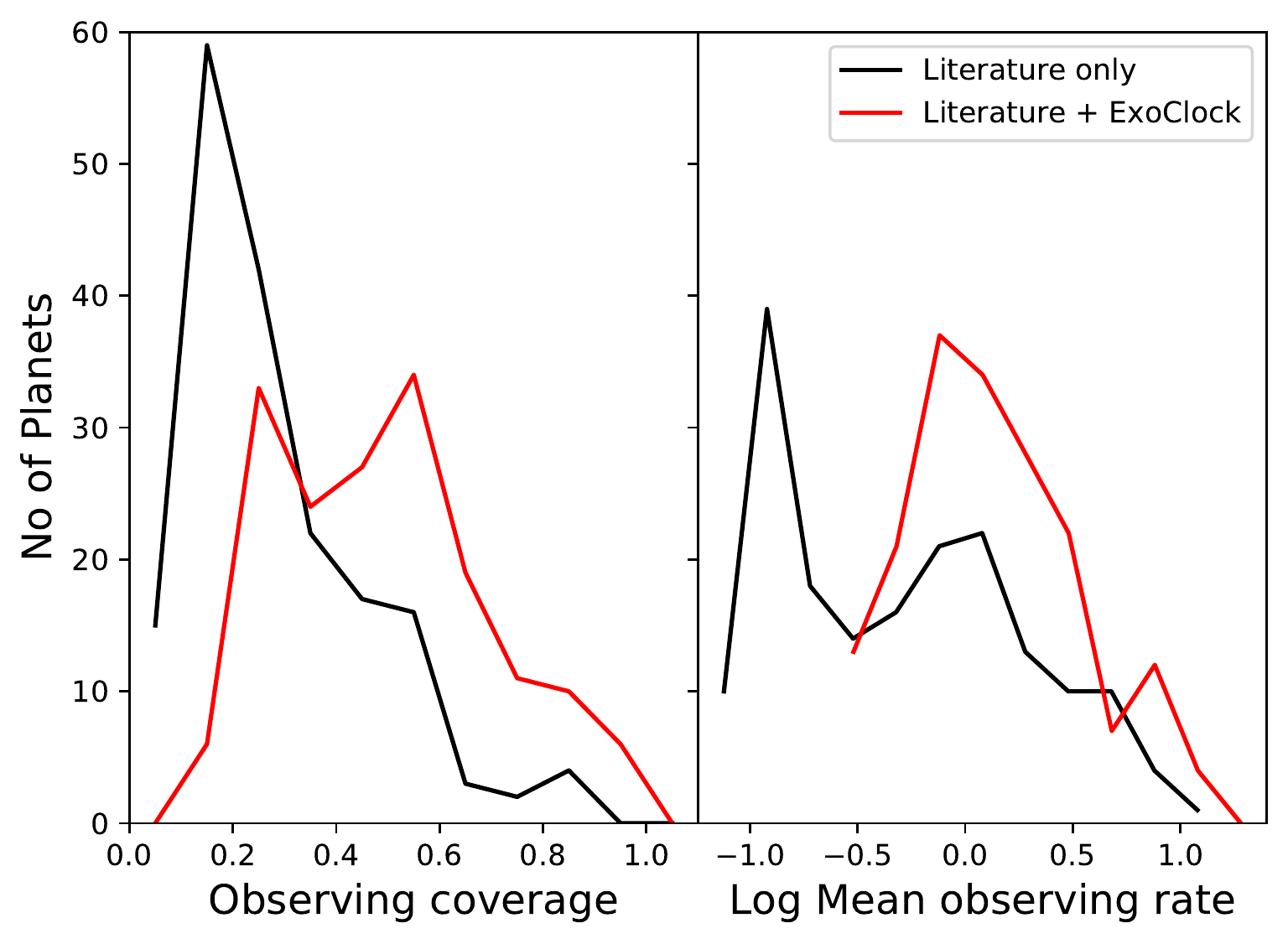}
\caption{Distribution of planets over the observing coverage (left) and the mean observing rate (right) for the literature data and for the combination of both literature and \exoclock\ data.}
\label{fig:rate_vs_coverage_distribution}
\end{figure}

Figure \ref{fig:rate_vs_coverage} shows the mean observing rate versus the observing coverage for all 180 planets, while figure \ref{fig:rate_vs_coverage_distribution} shows the distribution of planets over the two metrics. From both graphs, it is apparent that data collected as part of the \exoclock\ project have made a significant contribution to the follow-up of known planets, as the mean observing rate and the observing coverage have been doubled (or more) for 75\% of the planets.

Certainly, there are planets which have an adequate number of observations over the years (observing coverage $>$ 0.8), such as Qatar-1\,b. These planets have very reliable ephemerides and they can be considered as saturated. The large number of observations available for these targets is due to the fact that follow-up observations were focused on few specific targets. 

\begin{figure}
\centering
\includegraphics[width=\columnwidth]{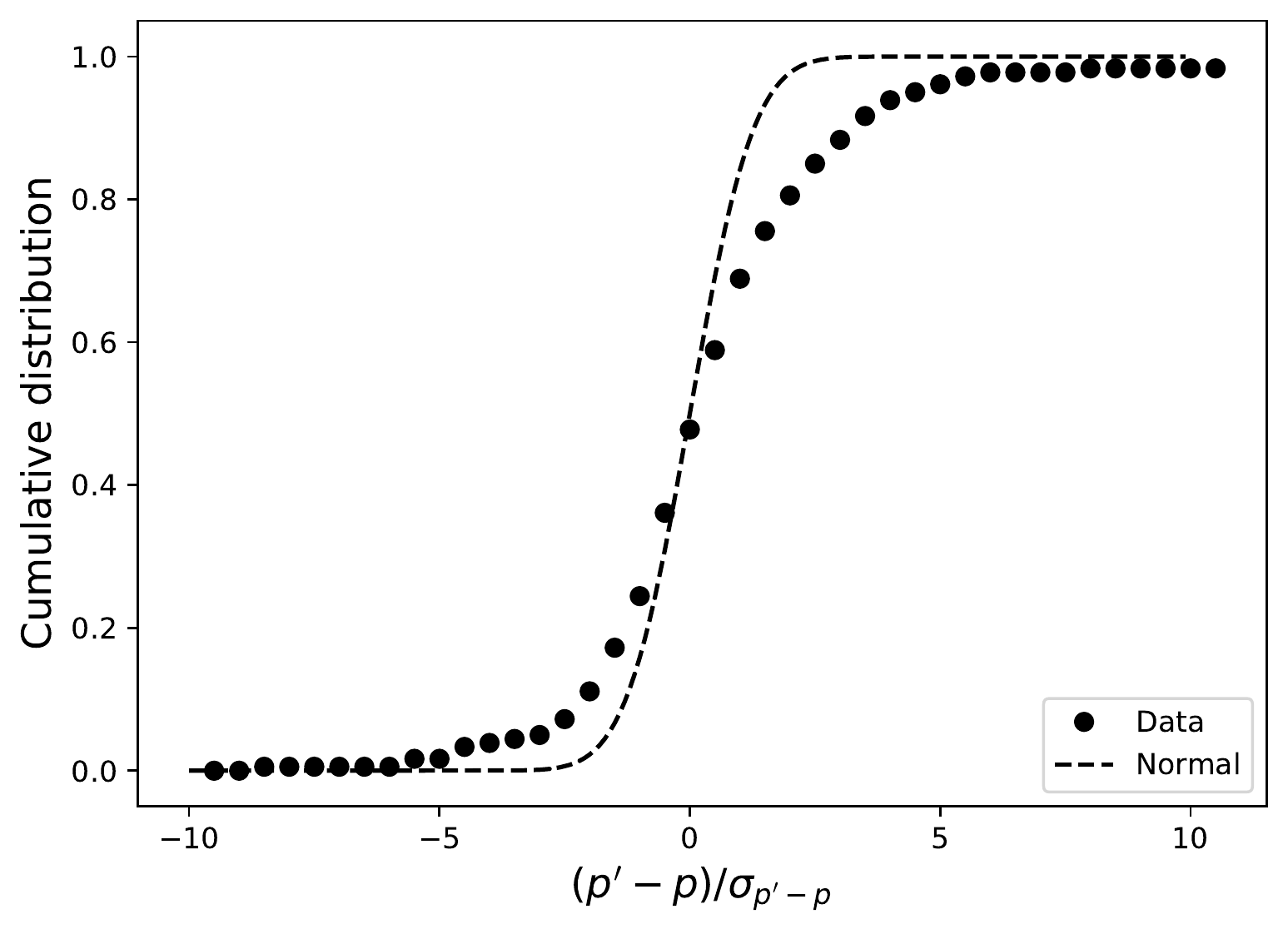}
\caption{Cumulative probability as a function of the drift in signal-to-noise-ratio and comparison with normal distribution.}
\label{fig:data_vs_gauss}
\end{figure}

However, the saturated planets constitute a small fraction (10\%) the total sample, while half of the planets still have an observing coverage lower than 0.5 and, therefore, they might show a drift in the future. Therefore, apart from the few planets which have been observed over a long time period and which have precise ephemerides, all the remaining planets require continuous monitoring in order to increase their time coverage as well. In addition, to maximise the time coverage for all the planets, in the near future we plan to incorporate more data from ETD, as well as the light-curves from Kepler, K2 and TESS in our database.

\subsection{General implications for follow-up observations of exoplanets}

The large-scale approach to the ephemerides refinement followed in this work enables a general study of the behaviour of the exoplanet ephemerides as a population. Here, we investigate the magnitude of the drifts between the predictions produced by the old and the new ephemerides (for the end of 2028) relative to their uncertainties. 

One would expect that the drifts are drawn from a normal distribution with a standard deviation equal to the prediction uncertainty. However, as we can see in Figure \ref{fig:data_vs_gauss} and also earlier in Figure \ref{fig:drifts}, the detected drifts are systematically greater than the prediction uncertainties. More specifically, 68\% of the detected drifts are within the $\pm2\sigma$ range, rather than the expected $\pm1\sigma$ range. Previous studies had indicated a similar but weaker effect based on a smaller sample of planets \citep{2019AnA...622A..81M}. This behaviour implies that the old ephemerides were mostly underestimating the prediction uncertainties. To understand if this behaviour is the result of biases in the calculation of the old ephemerides (most of which were calculated based on a small number of observations hence they could be biased more easily) or whether it is intrinsic to the follow-up strategy that is followed, it is necessary to repeat this type of evaluation regularly in the future.

\section{Conclusions}

The large-scale approach of this work is demonstrated to have significant implications for scheduling future observations for exoplanet characterisation studies. The examination of past literature mid-time values revealed that most planets lacked adequate observations while the focus of the studies was around a small number of planets. Observations provided by the \exoclock\ network doubled both the observing rate and coverage for half of the planets. Apart from the efficient organisation of the project where all available resources are utilised under the same scope, the findings confirmed that continuous monitoring of exoplanet ephemerides is crucial for follow-up studies, because a considerable fraction of the planets studied here (40\%) had highly uncertain or biased ephemerides. For this reason we plan to continue monitoring those planets, alongside other planets that have not been observed yet. A large number of observations ($\sim$600) for current and new targets has been submitted already to the \exoclock\ system and the results will be reported in a future study. All the results and the updated ephemerides are open to the wider exoplanet community to facilitate further research purposes in addition to \ariel.

\section*{Software and Data} 

Software used: Django, PyLightcurve \citep{Tsiaras2016B2016ascl.soft12018T}, ExoTETHyS \citep{Morello2020}, Astropy \citep{AstropyCollaboration2013}, emcee \citep{ForemanMackey2013}, Matplotlib \citep{Hunter2007}, Nestle, Numpy \citep{Harris2020}, SciPy \citep{Virtanen2020}.

All the data products and their descriptions can found through the OSF repository with DOI: \href{http://doi.org/10.17605/OSF.IO/WNA5E}{10.17605/OSF.IO/WNA5E} 

\section*{Acknowledgements}
This project has received funding from the European Research Council (ERC) under the European Union's Horizon 2020 research and innovation programme (grant agreement No 758892, ExoAI) and from the Science and Technology Funding Council (STFC) grants: ST/K502406/1, ST/P000282/1, ST/P002153/1, and ST/S002634/1. 

We would like to acknowledge the support provided by the administrators, designers, and developers of the ETD project and of the Czech Astronomical Society both to the ExoClock project but also to the efforts of the whole amateur community through its 10+ years of operation.

B. Edwards is the PI of the LCOGT Global Sky Partners project "ORBYTS: Refining Exoplanet Ephemerides" and thanks the LCOGT network and its coordinators for providing telescope access which resulted in data used in this publication. Additionally, B. Edwards acknowledges funding from STFC grant ST/T001836/1 and by the UKSA grant ST/V003380/1.

This work has made use of observations made by the MicroObservatory which is maintained and operated as an educational service by the Center for Astrophysics, Harvard \& Smithsonian as a project of NASA's Universe of Learning, supported by NASA Award \# NNX16AC65A.

L. V. Mugnai is supported by ASI grant n. 2018.22.HH.O. 

P. Pintr acknowledges the support by the project No.CZ.02.1.01/0.0/0.0/16\_026/0008390. 

A. Popowicz and K. Bernacki were responsible for automation of acquisition and processing of SUTO data and acknowledge grants: BK-225/RAu-11/2021, 02/140/SDU/10-22-02.

%REFERENCES
{\small
\bibliographystyle{aasjournals}
\bibliography{references,references_literature} 
}
%REFERENCES

\appendix
\newpage

\section{List of Private Observatories}

\begin{table}[h!]
\centering
\caption{List of private observatories beyond the list of affiliations.}
\label{tab:private_observatories}
\begin{tabular}{p{0.38\linewidth} p{0.48\linewidth}}
Observer(s) & Observatory \\ [0.3ex] \hline

Vikrant Kumar Agnihotri, Kuldip Vora & Cepheid Observatory, Rawatbhata, India \\ [-0.3ex]
David Bennett & Rickford Observatory, UK \\ [-0.3ex]
Paul Benni & Acton Sky Portal Observatory, Acton MA USA \\ [-0.3ex]
Emmanuel Besson & Observatoire des Loges, Buffières, France \\ [-0.3ex]
Leon Bewersdorff & Observatory Kipshoven, Germany \\ [-0.3ex]
Patrick Brandebourg & Observatoire du Guernet, Bretagne, France \\ [-0.3ex]
Stephen M. Brincat & Flarestar Observatory (MPC:171), San Gwann, Malta \\
\end{tabular}
\end{table}

\newpage

\section{Full list of updated ephemerides} \label{sec:updated_ephemerides}

\begin{longtable}{p{0.1\linewidth} | p{0.12\linewidth} | p{0.11\linewidth} | p{0.58\linewidth}}
\textbf{Planet} & \textbf{New T$_0$ (BJD$_\mathrm{TDB}$)} & \textbf{New Period (days)} & \begin{enumerate} \item \textbf{References for literature data used} \item \textbf{Reference for the initial ephemeris} \item \textbf{Reference for the transit parameters used} \item \textbf{Reference for the stellar parameters used} \end{enumerate} \\ [0.3ex] \hline

CoRoT-2b & 2457347.04314 $\pm$ 0.00012 & 1.74299700 $\pm$ 0.00000011 & \begin{enumerate} \item \cite{2008AnA...482L..21A,2019MNRAS.486.2290O,2010AJ....139...53R} \item \cite{2016AnA...595A..89B} \item \cite{2008AnA...482L..21A} \item \cite{2010AnA...517A..40C} \end{enumerate} \\ [0.3ex] \hline
GJ436b & 2454873.01582 $\pm$ 0.00004 & 2.64389751 $\pm$ 0.00000012 & \begin{enumerate} \item \cite{2014Natur.505...66K,2009AnA...507..481C,2016MNRAS.459..789T,2009ApJ...694.1559S,2007AnA...471L..51G,2008MNRAS.386.1644S,2007ApJ...667L.199D,2008ApJ...677L..59R,2008AnA...487L...5A,2007AnA...472L..13G,2008AnA...487L..25B,2011ApJ...731...16B} \item \cite{2014AnA...572A..73L} \item \cite{2011ApJ...735...27K} \item \cite{2008ApJ...677.1324T} \end{enumerate} \\ [0.3ex] \hline
GJ1214b & 2455881.579310 $\pm$ 0.000022 & 1.58040454 $\pm$ 0.00000004 & \begin{enumerate} \item \cite{2011ApJ...736...12B,2013ApJ...765..127F,2014AnA...565A...7C,2013AnA...549A..10H,2019RNAAS...3..123M} \item \cite{2014AnA...565A...7C} \item \cite{2012ApJ...747...35B} \item \cite{2009Natur.462..891C} \end{enumerate} \\ [0.3ex] \hline
HAT-P-1b & 2455801.77390 $\pm$ 0.00018 & 4.4652992 $\pm$ 0.0000004 & \begin{enumerate} \item \cite{2007AJ....134.1707W,2008ApJ...686..649J,2016MNRAS.459..789T} \item \cite{2014MNRAS.437...46N} \item \cite{2014MNRAS.437...46N} \item \cite{2008ApJ...677.1324T} \end{enumerate} \\
\end{longtable}

\end{document}